\documentclass[twocolumn,showpacs,preprintnumbers,amsmath,amssymb, prb]{revtex4}

\usepackage{graphicx}
\usepackage{dcolumn}
\usepackage{bm}

\begin{document}

\preprint{}

\title{Bound magnetic polarons in the very dilute regime}

\author{Yu.~G.~Kusrayev$^{1}$}
\author{K.~V.~Kavokin$^{1}$}
\author{G.~V.~Astakhov$^{2}$}\altaffiliation{Also at A. F. Ioffe Physico-Technical Institute,
RAS, 194021 St. Petersburg, Russia.}
\email[E-mail:]{astakhov@physik.uni-wuerzburg.de}
\author{W.~Ossau$^{2}$}
\author{L.~W.~Molenkamp$^{2}$}

\affiliation{$^{1}$A. F. Ioffe Physico-Technical Institute,
Russian Academy of Sciences, 194021 St. Petersburg, Russia \\
$^{2}$Physikalisches Institut (EP3), Universit\"{a}t W\"{u}rzburg,
97074 W\"{u}rzburg, Germany }

\date{\today}

\begin{abstract}
We study bound magnetic polarons (BMP) in a very diluted magnetic
semiconductor $\mathrm{Cd_{1-x}Mn_{x}Te}$ ($x < 0.01$) by means of
site selective spectroscopy. In zero magnetic field we detect a
broad and asymmetric band with a characteristic spectral width of
about 5~meV. When external magnetic fields are applied a new line
appears in the emission spectrum. Remarkably, the spectral width
of this line is reduced greatly down to $240$~$\mathrm{\mu eV}$.
We attribute such unusual behavior to the formation of BMP,
effected by sizable fluctuations of local magnetic moments. The
modifications of the optical spectra have been simulated by the
Monte-Carlo method and calculated within an approach considering
the nearest Mn ion. A quantitative agreement with the experiment
is achieved without use of fitting parameters. It is demonstrated
that the low-energy part of the emission spectra originates from
the energetic relaxation of a complex consisting of a hole and its
nearest Mn ion. It is also shown that the contribution to the
narrow line arises from the remote Mn ions.
\end{abstract}


\maketitle

\section{\label{sec1} INTRODUCTION }

An intriguing phenomenon in diluted magnetic semiconductors (DMS)
is the formation of bound magnetic polarons \cite{N1,N2}. The
bound magnetic polaron (BMP) is a local ordering of magnetic
moments induced by the exchange interaction with a localized
carrier. Most BMP studies, both experimental and theoretical, have
been performed so far in a regime of the mean-field approach (when
$N_{\mathrm{Mn}} a_B^3 \gg 1$) \cite{N3,N4,N5,N6,N7,N8}. Here,
$N_{\mathrm{Mn}} = x N_0$ is the concentration of magnetic ions,
$N_0$ is the concentration of cations, and $a_B$ is the
localization radius. This approach implies that the localized
carrier interacts with a infinite number of $\mathrm{Mn}^{2+}$
ions, and hence it is typically applicable for high and moderate
manganese concentrations ($x > 0.01$). At present systems with a
countable number of magnetic moments attract growing interest
\cite{N_QD}. This corresponds to the very dilute regime
$N_{\mathrm{Mn}} a_B^3 \ll 1$, i.e. when on average one carrier
interacts with one (or few) Mn ion(s). Clearly, at this condition
the mean-field approach is not an appropriate model.

Up to now bound magnetic polarons have not been observed in a very
dilute regime. This is because their observation requires the
polaron formation time $\tau_F$  to be shorter than the life time
$\tau$ of nonequilibrium carriers, $\tau_F < \tau$. In DMS with $x
\gtrsim 0.1$ this condition is fulfilled. However, in samples with
lower Mn concentrations $x \leq 0.01$ the polaron formation time
is of the order $\tau_F \sim 10^{-4} \div 10^{-6}$~s
\cite{N9,N10}. Therefore, the BMP formation process is interrupted
by exciton recombination (with characteristic time $\tau \sim
10^{-9}$~s) \cite{N11}.

In this paper we study BMP in bulk $\mathrm{Cd_{1-x}Mn_{x}Te}$
samples with very low Mn concentration ($x < 0.01$), i.e. at the
condition $N_{\mathrm{Mn}} a_B^3 \ll 1$. In order to avoid the
above problem of BMP detection in the very dilute regime we use
resonant (selective) excitation of donor-acceptor (DA) pairs with
transition energies \cite{N4}
\begin{equation}
E_{DA} = E_g - E_D - E_A - e^2 / \epsilon \rho  \,. \label{Eq_DA}
\end{equation}
Here, $E_D$ and $E_A$ are the binding energies of donors and
acceptors respectively, $E_g$ is the band gap and $\epsilon$ is
the dielectric constant in $\mathrm{Cd_{1-x}Mn_{x}Te}$. As follows
from Eq.~(\ref{Eq_DA}) by changing the excitation energy $E_{ex} =
E_{DA}$ one can selectively excite pairs with a given DA
separation $\rho$. Because of the weak overlap [$\sim \exp(- \rho
/ a_B)$] of the donor and acceptor wavefunctions, the pairs with
large $\rho$ are characterized by a long recombination time $\tau
> \tau_F$. As a result the bound magnetic polaron forms before the
recombination occurs. Without polaronic effects such resonantly
excited pairs should manifest themselves in emission spectra as a
narrow line of width within either the inverse recombination time
($\hbar / \tau$) or the laser line width.

\section{\label{sec2} EXPERIMENT}

The bulk $\mathrm{Cd_{1-x}Mn_{x}Te}$ samples with Mn
concentrations $x = 0.003, \, 0.005, \, 0.02$ have been grown by
the Bridgman technique. The crystals are nominally undoped. The
concentration of residual impurities (both donors and acceptors)
is about $10^{16}$~$\mathrm{cm}^{-3}$, which is much lower than
the Mn concentration $N_{\mathrm{Mn}}$. As grown crystals are
usually p-type due to excess of Te or Cd-vacancies. Note, for
acceptors with the Bohr radius $a_B = 10$~{\AA} and $x = 0.005$
($N_{\mathrm{Mn}} = 7 \times 10^{19}$~ $\mathrm{cm}^{-3}$) the
relationship is $N_{\mathrm{Mn}} a_B^3 = 0.07$. In the experiments
we used $4 \times 4 \times 0.3$~mm pieces cleaved out along the
(110) plane from the massive monocrystals. All measurements were
carried out at a temperature $T = 1.6$~K. External magnetic fields
were applied either perpendicular to the sample plane (Faraday
geometry) or in the sample plane (Voigt geometry). The
photoluminescence (PL) was excited by a He-Ne laser or a tunable
Ti-sapphire laser pumped by an Ar-ion laser. The excitation
density was about 5~$\mathrm{W / cm^2}$. The laser beam was
directed onto the samples at the angle close to the normal (axis
[110]), and the emission was registered in the backscattering
geometry. A polarizer (in combination with a quater-wave plate)
was (were) used to linearly (circularly) polarize the excitation.
The degree of PL polarization was detected with use of a
photo-elastic modulator and a two-channel photon counter.

\begin{figure}[btp]
\includegraphics[width=.41\textwidth]{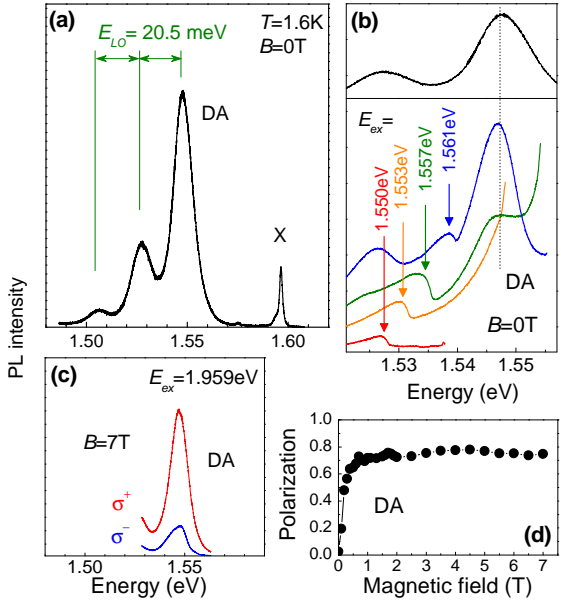}
\caption{(Color online) (a) Zero-field PL spectrum of the
$\mathrm{Cd_{0.995}Mn_{0.005}Te}$ sample excited by a He-Ne laser
($E_{ex} = 1.959$~eV). (b) PL spectra under quasiresonant
excitation, the excitation energies ($E_{ex}$) being shown in the
panel. Arrows indicate emission bands which are related to the BMP
formation. PL in the same spectral range but under nonresonant
excitation [as in (a)] is given in the upper part for comparison.
(c) PL spectra of DA pairs recorded in $\sigma^{+}$ and
$\sigma^{-}$ circular polarizations in an external magnetic field
$B = 7$~T applied in Faraday geometry. The excitation conditions
are the same as in (a). (c) Degree of circular polarization vs
magnetic field detected at the DA band. } \label{fig1}
\end{figure}

We found qualitatively very similar behavior in all samples
studied. In the following we present experimental results for one
piece cleaved out from the bulk $\mathrm{Cd_{0.995}Mn_{0.005}Te}$
sample. Figure~\ref{fig1}(a) shows the PL spectrum excited above
the CdTe band gap $E_g$ (He-Ne laser with $E_{ex} = 1.959$~eV).
The high-energy line (X) is attributed to the bound exciton. At
low-energy part of the PL spectrum we observe a broad band
associated with recombination of DA pairs and its phonon replicas.
These replicas are separated by 20.5~meV, which corresponds to the
LO-phonon energy in CdTe \cite{N12}. Application of external
magnetic fields in Faraday geometry results in in the circularly
polarized emission already in low fields [Fig.~\ref{fig1}(c)].
Such a behavior is typical for magnetic semiconductors.
Figure~\ref{fig1}(d) demonstrates the degree of circular
polarization vs magnetic field detected at the DA band. It
saturates rapidly on a 75\% level.

\begin{figure}[tbp]
\includegraphics[width=.43\textwidth]{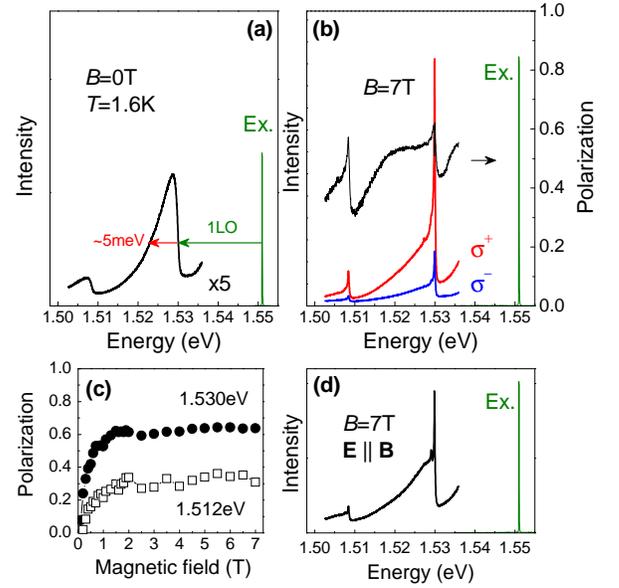}
\caption{(Color online) (a) BMP spectrum obtained under resonant
excitation (Ex.) to the DA band ($E_{ex} = 1.551$~eV) in zero
magnetic field. (b) BMP and BMP-polarization spectra obtained
under the same excitation conditions as in (a) but in magnetic
field $B = 7$~T applied in Faraday geometry. (c) degree of
circular polarization vs magnetic field detected at different
energies. (d) BMP spectrum obtained in magnetic field $B = 7$~T
applied in Voigt geometry when the polarization axis of the
excitation ($E$) is parallel to the field direction ($\mathbf{E}
|| \mathbf{B} $). } \label{fig2}
\end{figure}

In order to study bound magnetic polarons we use the site
selective spectroscopy \cite{N13}. We resonantly excite DA pairs
and record BMP spectra shifted towards lower energies with respect
to the excitation energy $E_{ex}$ (the Stokes shift). A set of PL
spectra for different $E_{ex}$ is presented in Fig.~\ref{fig1}(b).
A new emission band (indicated by arrows), being observed on the
top of the DA photoluminescence, clearly follows the laser line
which is a fingerprint of the BMP formation. The most pronounced
results have been obtained for $E_{ex} = 1.551$~eV, as also shown
in Fig.~\ref{fig2}. In order to avoid scattered light from the
laser we detect such BMP spectra at the first phonon replica.

In zero magnetic field the BMP spectrum represents itself as an
asymmetric band with the 5-meV-tail towards low energies
[Fig.~\ref{fig2}(a)]. Our estimations (given below) show that this
value is in agreement with the BMP energy as well as with its
root-mean-square fluctuations. Such a behavior is an essential
deviation from the case of high Mn concentrations
($N_{\mathrm{Mn}} a_B^3 \gg 1$), where the Stokes shift is larger
than the line width, giving rise to a distinct emission line
shifted from the excitation by the BMP energy \cite{N13}.

\begin{figure}[tbp]
\includegraphics[width=.43\textwidth]{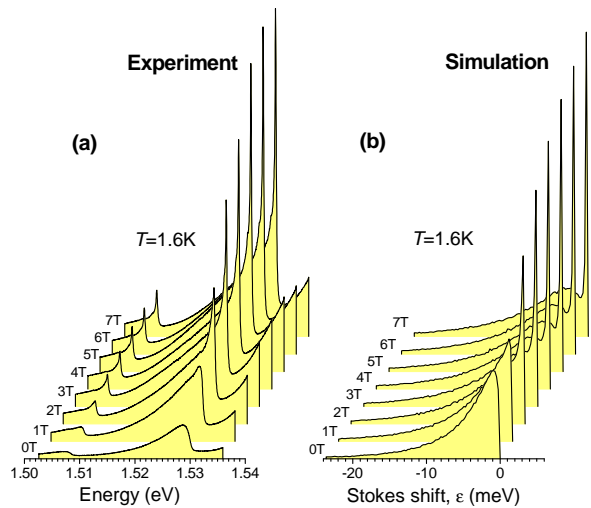}
\caption{(Color online) Evolution of BMP spectra with growing
magnetic field: (a) experimant and (b) the Monte-Carlo simulation.
Note, in (b) no fitting parameters are used. The PL spectra are
shifted for clarity. } \label{fig3}
\end{figure}

The most drastic changes, however, occur in external magnetic
fields applied in Faraday geometry [Fig.~\ref{fig3}(a)]. A new
line appears at the resonant energy (i.e., there is no Stokes
shift at all) \cite{comment}. With growing magnetic field the
amplitude of this line increases and the width reduces greatly. In
strong enough magnetic fields ($B > 3$~T) the width saturates on a
$240$-$\mathrm{\mu eV}$-level [see also Fig.~\ref{fig2}(b)]. The
circular polarization of this line is about 65\%, and its
dependence on the magnetic field [Fig.~\ref{fig2}(c)] follows that
for the DA emission under nonresonant excitation
[Fig.~\ref{fig1}(d)]. One more lineament of this new line: Its
intensity reduces under excitation by the light being linearly
polarized along the magnetic field direction applied in Voigt
geometry [see Fig.~\ref{fig2}(d)]. However, a deviation of the
field direction from the sample plane or the polarization axis
from the field direction restores the amplitude of this line.

\begin{figure}[btp]
\includegraphics[width=.41\textwidth]{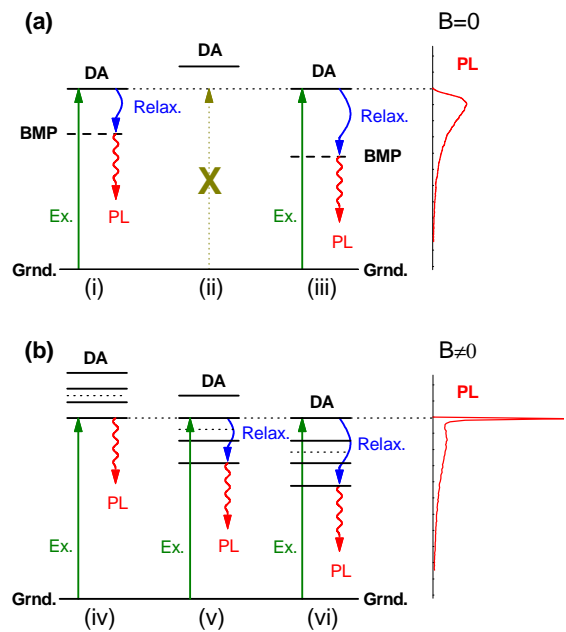}
\caption{(Color online) Formation of BMP spectra under selective
excitation (Ex.) in (a) zero and (b) strong magnetic fields. The
roman numerals in brackets enumerate different DA pairs. }
\label{fig4}
\end{figure}

As we show in Sec.~\ref{sec3}, the evolution of PL spectra
presented in Fig.~\ref{fig3}(a) cannot be described in terms of
the mean-field theory, and hence this requires an alternative
approach. First, we discuss the asymmetric band observed in zero
magnetic field. Resonant excitation generates DA pairs with a
certain energy $E_{DA} = E_{ex}$ depending on donor-acceptor
separation $\rho$ [pairs (i) and (iii) in Fig.~\ref{fig4}(a)]. DA
pairs with transition energies different from the excitation
energy $E_{DA} \neq E_{exc}$ are not excited [pair (ii) in
Fig.~\ref{fig4}(a)]. The observation of the asymmetric band in
Fig.~\ref{fig2}(a) we attribute to the BMP formation accompanied
by relaxation to the ground state. The dominant contribution to
the polaron shift arises from the hole. This is because the
absolute value of the $p$-$d$ exchange constant is larger than the
absolute value of the $s$-$d$ exchange, and the Bohr radius of the
acceptor is smaller than that of the donor \cite{N1,N2}. In this
approach the line width is explained by the dispersion of the
polaron shifts induced by static fluctuations of Mn concentration.
For instance, pair (iii) in Fig.~\ref{fig4}(a) shows a larger
polaron shift than pair (i). Because on average less than one Mn
falls into the volume restricted by the acceptor Bohr radius, the
root-mean-square fluctuations appear to be comparable with the
polaron shift. Its value can be estimated from the line width in
Fig.~\ref{fig2}(a), $\Gamma \approx 5$~meV.

Strong enough magnetic fields suppress the polaron formation
\cite{N14}, and one would expect the disappearance of the
low-energy tail. However, our experiments show the opposite
behavior: This tail preserves and a new narrow intense line
appears. We suggest the following explanation. Acceptor (donor)
states are split in an external magnetic field. The photon
absorption by the fixed DA pair leads to the population of one of
the Zeeman sublevels [Fig.~\ref{fig4}(b)]. Relaxation from the
higher-lying sublevels to the lowest sublevel [pairs (v) and (vi)
in Fig.~\ref{fig4}(b)] results in the low-energy tail. The
characteristic width of this tail does not change with magnetic
fields because the Zeeman splittings disperse in the same manner
as the polaron shifts, i.e. due to the static fluctuations of the
Mn concentration. In case when the lowest Zeeman sublevel of DA
pairs is excited [pair (iv) in Fig.~\ref{fig4}(b)] a further
relaxation is not possible, and the narrow line is observed at the
excitation energy. With growing magnetic fields the width of this
line should decrease and saturate in fields when the magnetization
of magnetic ions saturates ($B > 3$~T).

It is now clear why the narrow peak is suppressed when the
polarization axis of the excitation light is exact parallel to the
magnetic field direction: Due to optical selection rules the
excitation to the ground state [pair (iv) in Fig.~\ref{fig4}(b)]
is forbidden in this configuration. However, even a small
deviation from this geometry restores the perpendicular-to-field
component of the polarization vector, and the peak rises up.

\section{\label{sec3} THEORY}

In order to calculate the PL spectra under selective excitation
one needs to find the distribution function of the Stokes shift
induced by the polaron formation in zero magnetic field. On the
other hand, in case of strong magnetic fields the distribution
function is determined by the Zeeman splitting dispersion. 
The hole (electron) spins interact with Mn ions via the $p$-$d$
($s$-$d$) exchange interaction. In case of localized carriers this
interaction can be written as
\begin{eqnarray}
\mathcal{\widehat{H}}_{sd} = \alpha \sum_{i} \overrightarrow{s_e}
\cdot \overrightarrow{S_i}  \, \Psi_e^2 (\overrightarrow{r_i}) \,,
\label{Eq1}
\\
\mathcal{\widehat{H}}_{pd} = \frac{\beta}{3} \sum_{i}
\overrightarrow{J} \cdot \overrightarrow{S_i}  \, \Psi_h^2
(\overrightarrow{r_i}) \,, \label{Eq2}
\end{eqnarray}
where $\overrightarrow{s_e}$ and $\overrightarrow{J}$ are the
electron and hole spin operators, $\Psi_e$ and $\Psi_h$ are their
wavefunctions, respectively. $\overrightarrow{S_i}$ is the spin
operator of the $i$-th Mn ion, and $\overrightarrow{r_i}$ is a
radius-vector assigning Mn location in the crystal. The exchange
constants in $\mathrm{Cd_{1-x}Mn_{x}Te}$ are $\alpha =
14850$~$\mathrm{meV \, \AA^3}$ and $\beta = 59400$~$\mathrm{meV \,
\AA^3}$. The donor and acceptor Bohr radii are equal approximately
60~{\AA} and 10~{\AA}, respectively. This implies that the spin
splitting of the Mn ion removed by the Bohr radius from the
acceptor (donor) is 1.28~meV (1.5~$\mathrm{\mu eV}$). In other
words, at a temperature $T = 1.6$~K ($k_B T = 0.14$~meV) the
electron contribution can be neglected, and all Mn spins in the
vicinity of the acceptor are aligned along the exchange field of
the hole. The hole exchange energy in the ground state is $E_h =
\frac{5}{4} \beta \sum_{i} \Psi_h^2 (\overrightarrow{r_i})$. Its
mean value averaged over all Mn configurations is $\langle E_h
\rangle = \frac{5}{4} \beta N_0 x$, where $N_0 =
0.015$~$\mathrm{\AA^{-3}}$ is a concentration of cations. For
$x=0.5\%$ it is equal $\langle E_h \rangle = 5.0$~meV. The
root-mean-square fluctuation $\delta E_h = \sqrt{\langle E_h^2
\rangle - \langle E_h \rangle^2}$ is equal $\delta E_h = 5.9$~meV.
And the mean value of the hole energy in the fluctuating fields of
nonpolarized Mn ions $\delta E_{hf} = \frac{1}{2} \beta \sqrt{N_0
x \frac{S (S+1)}{3} \int \Psi_h^3 d^3r }$ is equal $\delta E_{hf}
\approx 5.5$~meV. It follows from these estimations that the
fluctuations of the exchange energy and its mean value are nearly
the same. Therefore the Gaussian statistics and, as consequence,
the mean-field approach is not applicable.

\begin{figure}[btp]
\includegraphics[width=.45\textwidth]{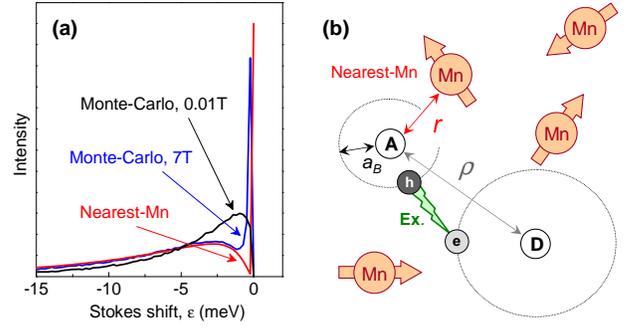}
\caption{(Color online) (a) Calculations of the PL spectra by the
Monte-Carlo method and within the approach considering the
nearest-Mn ion. Note, the nearest-Mn approach gives the same
result for zero and strong magnetic fields. (b) Random
distribution of acceptors (A), donors (D) and Mn ions. }
\label{fig5}
\end{figure}

In case of low Mn concentrations it is difficult to obtain an
analytical solution for the Stokes shift distribution function.
Therefore, in order to calculate the PL spectra under selective
excitation we use numerical computing based on the Monte-Carlo
method. First, a random distribution of magnetic ions in a sphere
with a radius equal to ten acceptor Bohr radii $10 a_B$ is
generated [see also Fig.~\ref{fig5}(b)] \cite{N15}. Each Mn ion
has a random spin orientation in accordance with the equilibrium
distribution for a given temperature and magnetic field strength.
Then, the total exchange field acting on the hole bound to the
acceptor placed in the center of the sphere is calculated. After
that, the eigenstates of the hole in this total exchange field and
the transition probabilities on these states are evaluated.
Polaron energies are determined for each generated Mn distribution
by adding the exchange filed of the hole acting on Mn ions to the
external magnetic field. The Stokes shift is calculated as an
energy difference between the initial and final states. Finally
the distribution function of the Stokes shift is obtained after
summation over  50000 realizations.

The calculated PL spectra for different magnetic fields are shown
in Fig.~\ref{fig3}(b). Not only qualitative but also quantitative
agreement with experimental data [compare with Fig.~\ref{fig3}(a)]
is achieved. Note, no fitting parameters are used in the
calculation.

Because of low Mn-ion count in the vicinity of the acceptor (on
average there are 0.3 Mn ions inside the sphere of radius $a_B$)
it is natural to assume that the nearest Mn ions give dominant
contributions to the Stokes shifts [Fig.~\ref{fig5}(b)]. We now
consider this approach in details. In an external magnetic field
the hole states are split in four sublevels characterized by the
different projection $J_z$ of the hole angular momentum on the
field direction. One of them with $J_z = 3/2$ is a ground state.
The contribution to the PL spectra from this state is a delta
function $\delta (\varepsilon)$ (if neglecting native homogeneous
broadening) as shown in  Fig.~\ref{fig4}(b) [pair (iv)].
Excitation in one of three other states is followed by relaxation
to the ground state [pairs (v) and (vi) in Fig.~\ref{fig4}(b)].
Therefore, the shift of the emission energy with respect to the
excitation energy is described by the distribution function of the
nearest (to the hole) Mn ion, $F_{NN} (\varepsilon)$. Then, under
unpolarized excitation the distribution function $F(\varepsilon)$
of the Skokes shift $\varepsilon$ is written as
\begin{eqnarray}
F (\varepsilon)= W_{3/2} \delta (\varepsilon) + W_{1/2} F_{NN} (3
\varepsilon / 2 \varepsilon_0) +
\,\,\,\,\,\,\,\,\,\,\,\,\,\,\,\,\,\,\,\,\,\,
\nonumber\\
\,\,\,\,\,\,\,\,\, +  W_{1/2} F_{NN} (3 \varepsilon / 4
\varepsilon_0) + W_{3/2} F_{NN} ( \varepsilon / 2 \varepsilon_0)
\,, \label{Eq3}
\end{eqnarray}
where $W_{3/2} = 3/4$ and $W_{1/2} = 1/4$ are the probabilities to
excite the hole with $J_z = \pm 3/2$ and $J_z = \pm 1/2$,
respectively. And the characteristic exchange energy is
$\varepsilon_0 = \frac{5}{4} \beta \Psi_h^2 (0) = 23.6$~meV. The
probability $dP_{NN} (r)$ that the nearest Mn ion is located in
the spherical layer ($r$,$r + dr$) [see Fig.~\ref{fig5}(b)] is
given by the probability to find one Mn ion in this spherical
layer being equal to $x N_0 4 \pi r^2 dr$, and the probability
that there are no Mn ions inside the sphere of radius $r$ being
equal to $P_0 (r)$. Here, $P_0 (r)$ is the Poisson distribution
$P_n (\overline{n}) = \frac{\overline{n}^n}{n!} \exp (-
\overline{n})$ with $n = 0$ and $\overline{n} = \frac{4}{3} \pi
r^3 x N_0$. Then one writes
\begin{equation}
dP_{NN} (r)= 4 \pi r^2 x N_0 \, \exp \left( - 4 \pi r^3 x N_0 / 3
\right) dr \,. \label{Eq4}
\end{equation}
Recall, the energy of the exchange interaction between the hole
and Mn ion depends on a distance $r$ as $\Psi_h^2 (r)$, i.e.
$\varepsilon (r) = \varepsilon_0 \exp (-2 r / a_B)$ and $dr = -
a_B / (2 \varepsilon) d \epsilon$. Finally, one obtains the
distribution function vs energy $\varepsilon$ as $F_{NN} = dP_{NN}
/ d \varepsilon$
\begin{equation}
F_{NN} (\varepsilon / \varepsilon_0)= \frac{3 \gamma}{2
\varepsilon} \left( \ln \frac{\varepsilon}{\varepsilon_0} \right)
^2 \exp \left[ \gamma \left( \ln \frac{\varepsilon}{\varepsilon_0}
\right) ^3 \right] \,, \label{Eq5}
\end{equation}
with $\gamma = (\pi / 6) x N_0 a_B^3$.

Figure~\ref{fig5}(a) shows that the nearest-Mn approach describes
well the PL spectra in high magnetic fields. However, the
agreement becomes worse for small $\varepsilon$ in zero magnetic
field, while the high-energy tail is still described quite well.
The reason is that for the complex 'hole + nearest-Mn ion' the
relaxation is not possible at the excitation to the ground state,
and the Stokes shift is absent. This results in a
$\delta$-function peak appearing in zero as well as in high
magnetic fields [first term in Eq.~(\ref{Eq3})]. In real the
contribution to the Stokes shift for $\varepsilon \ll
\varepsilon_0$ arises from several distant ($r \gg a_B$) Mn ions.
In weak magnetic fields the spins of these distant ions are
disordered, and their orientation induced by the exchange field of
the hole results in the energy lowering. In strong magnetic fields
the spins are already oriented at the moment of excitation,
further relaxation does not occur, which manifests itself as the
appearance of the narrow peak at $\varepsilon = 0$.

\section{\label{sec5} CONCLUSIONS}

With use of the site selective spectroscopy of residual DA pairs
in bulk $\mathrm{Cd_{1-x}Mn_{x}Te}$ samples we are able to observe
the formation of bound magnetic polarons in a very dilute ($x <
0.01$) regime. Under selective excitation the PL spectra show an
asymmetric band with low-energy tail and the additional very
narrow peak inflaming in external magnetic fields. This behavior
cannot be described in terms of the mean-field approach,
frequently applied to diluted magnetic semiconductors (DMS). Our
findings suggest that the low-energy tail is formed due to the
exchange interaction of the hole (bound to the acceptor) with the
nearest Mn ion, and is described by the Poisson distribution.
Unexpectedly, the dominant contribution to the narrow peak arises
from the interaction with several Mn ions remote from the hole.

\begin{acknowledgments}

This research was supported by the DFG (436 RUS 113/843 and SPP
1285) as well as the RFBR. We thank prof.~R.~R.~Ga{\l}\c{a}zka
(Institute of Physics, Warsaw, Poland) for providing us with high
quality samples.

\end{acknowledgments}



\begin{references}
\bibitem{comment} In continuous wave (cw) experiments it is not possible
to separate the emission from the laser light at the energy of
excitation. But this can be easily done at the phonon replicas of
PL lines, where the contribution of scattered light is negligible.
\bibitem{N15} Note in Ref.~\onlinecite{N4}, in order to calculate BMP
energies, the magnetic ions in a sphere of radius $R \sim 1.5 a_B$
were considered.
\end{references}
\end{document}